\input{aipcheck}

\documentclass[
    ,final            
  ]
  {aipproc}

\layoutstyle{8x11single}
\newcommand{\grado}{^{\circ}}

\begin{document}

\title{Dark Matter annihilation in Draco: new considerations of the expected gamma flux \footnote{Poster presented at the First GLAST Symposium, Stanford University, USA, 5-8 February 2007}}

\classification{95.35.+d; 95.55.Ka; 95.85.Pw; 98.35.Gi; 98.52.Wz}
\keywords      {cosmology: dark matter --- galaxies: dwarf --- gamma-rays: theory}

\author{Miguel A. S\'anchez-Conde}{
  address={Instituto de Astrofisica de Andalucia (CSIC), E-18008,
Granada, Spain} }

\begin{abstract}
A new estimation of the $\gamma$-ray flux that we expect to detect
from SUSY dark matter annihilation from the Draco dSph is presented
using the DM density profiles compatible with the latest
observations. This calculation takes also into account the important
effect of the Point Spread Function (PSF) of the telescope. We show
that this effect is crucial in the way we will observe and interpret a
possible signal detection. Finally, we discuss the prospects to detect
a possible gamma signal from Draco for MAGIC and GLAST.
\end{abstract}

\maketitle


\section{Introduction}
The Draco galaxy, a satellite of the Milky Way, represents one of the
  best suitable candidates to search for DM outside our galaxy \citep{evans}, since
  it is near (80 kpc) and it has probably more observational
  constraints than any other known DM dominated system. This fact
  becomes crucial when we want to make realistic predictions of the
  expected observed $\gamma$-ray flux due to DM annihilation.

The expected total number of continuum $\gamma$-ray photons received
per unit time and per unit area, from a circular aperture on the sky
of width $\sigma_{\rm t}$ (which represents the resolution of the
telescope) observing at a given direction $\Psi_0$ relative to the
centre of the dark matter halo is given by:

\begin{equation}
F(E>E_{\rm th})=\frac{1}{4\pi} {f_{SUSY}} \cdot U(\Psi_0), \qquad \textrm{with} \qquad f_{SUSY}= \frac{N_{\gamma} \left<\sigma v\right>}{2 m_\chi^2}, \qquad U(\Psi_0)=\int J(\Psi)B(\Omega)d\Omega
\label{eq1}
\end{equation}

\noindent where the factor $f_{SUSY}$ encloses all the particle physics, and the factor $U(\Psi_0)$ involves all the astrophysical properties such as the dark matter distribution, geometry considerations and telescope performances like the PSF, this one directly related to the angular resolution (for a detailed explanation of each of these terms, see e.g.~\citep{prada}).

\section{Draco $\gamma$-ray flux profiles and the effect of the PSF}

\begin{figure}
  \begin{minipage}[t]{0.5\linewidth} \centering
    \includegraphics[height=.17\textheight,width=.8\textwidth]{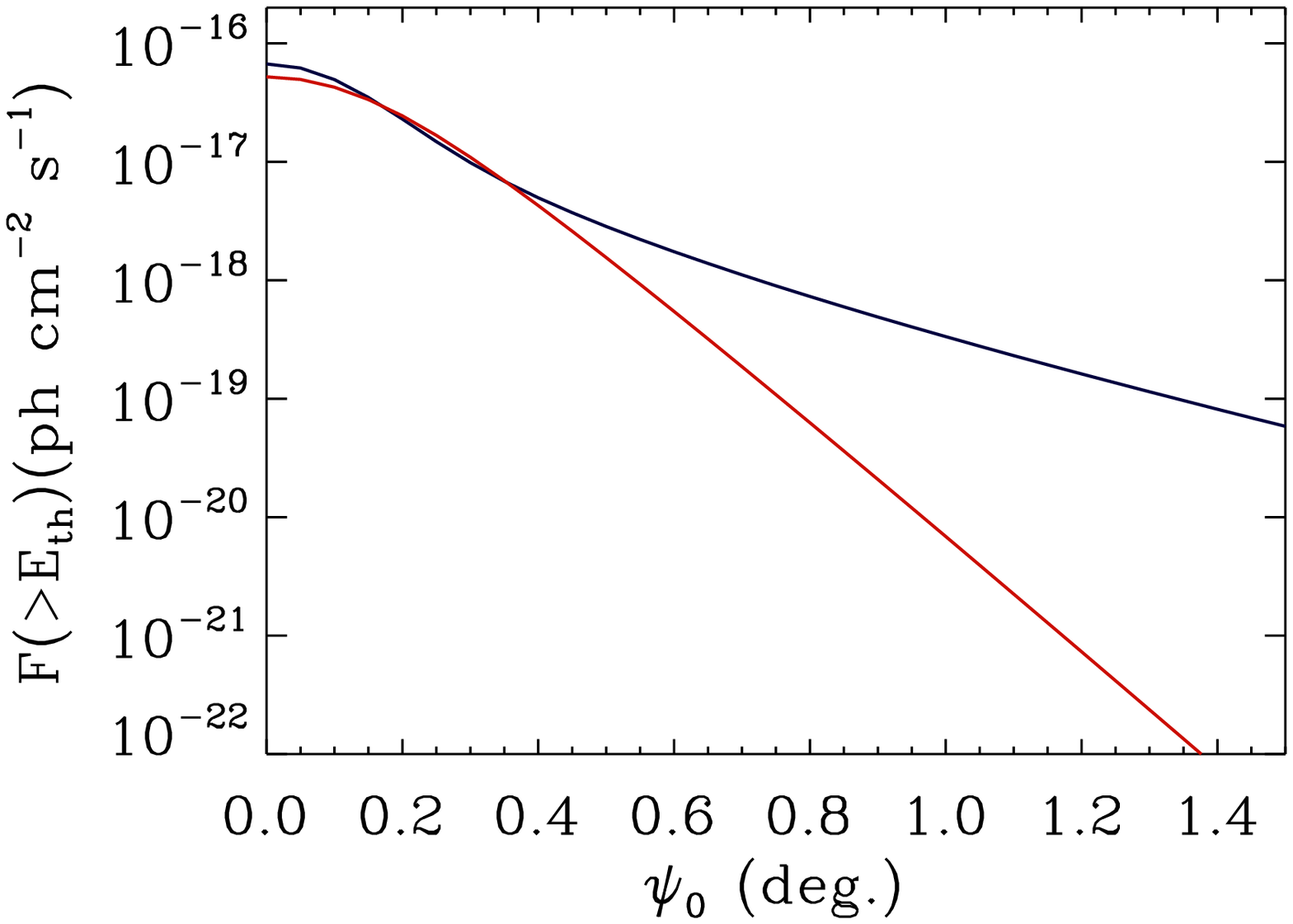}
    \end{minipage} 
    \begin{minipage}[t]{0.5\linewidth} \centering
    \includegraphics[height=.17\textheight,width=.8\textwidth]{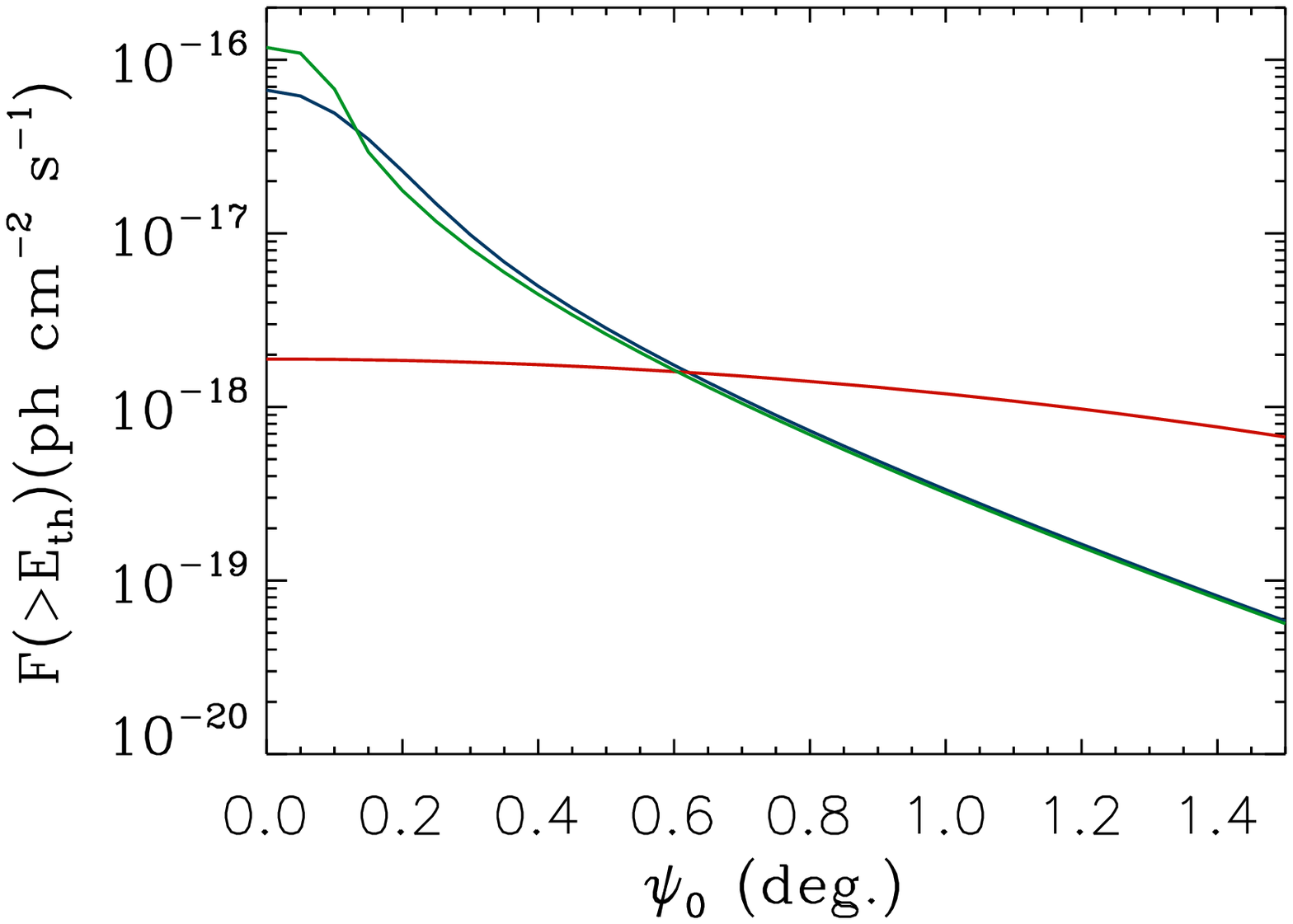}
    \end{minipage} \caption{Left panel: Draco flux predictions for the
    core (red line) and cusp (blue line) DM density profiles, computed
    using a PSF= 0.1$\grado$. A value of $f_{SUSY}= 10^{-33}
    GeV^{-2}cm^3s^{-1}$ was used (the most
    optimistic case given by particle physics at 100 GeV). Right panel:
    Flux predictions for the cusp density profile and using a PSF=0.1$\grado$ (blue line), PSF=1$\grado$ (red) and without PSF (green).}
    \label{fig1}
\end{figure}

 We assumed that the dark matter distribution in Draco can be
approximated by the formula $\rho_{\rm d}(r) = C r^{-\alpha} {\rm exp}
(-\frac{r}{r_{\rm b}})$ proposed by \citep{kmmds}, which was found to
fit the density distribution of a simulated dwarf dark matter halo
stripped during its evolution in the potential of a giant galaxy. Here
we will consider two cases, the profile with a cusp $\alpha=1$ and a
core $\alpha=0$. The calculated $\gamma$-ray flux profiles are shown
in left panel of Fig.\ref{fig1}.  For both cusp and core DM density
profiles, the flux values should be very similar for the inner region
of the dwarf, where signal detection would be easier. It is also
necessary to take into account the role of the PSF in the
calculations. Its effect on gamma ray flux profiles, usually
neglected, may be crucial to correctly interpret a possible signal in
the telescope, as it can be clearly seen in right panel of
Fig. \ref{fig1}.


\section{Detection prospects}

\begin{figure}
  \includegraphics[height=.18\textheight,width=.4\textwidth]{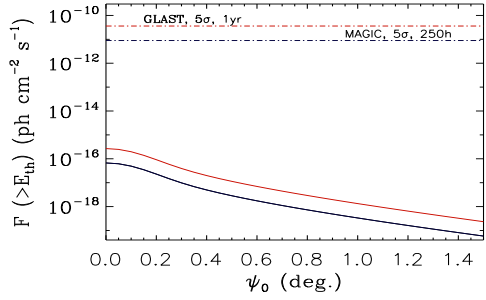}
  \caption{Draco flux profile detection prospects for GLAST (red
  lines) and MAGIC (blue lines), and for the cusp density profile
  using a PSF=0.1$\grado$. Values of $f_{SUSY}= 10^{-33}
  GeV^{-2}cm^3s^{-1}$ at 100 GeV and $f_{SUSY}= 4~10^{-33}
  GeV^{-2}cm^3s^{-1}$ at 10 GeV were used for MAGIC and GLAST
  respectively (the most optimistic scenarios given by
  particle physics at those energies.}  \label{fig2}
\end{figure}

We carried out some calculations concerning the possibility to detect
a $\gamma$-ray signal coming from DM annihilation in Draco for MAGIC
and GLAST (Fig.\ref{fig2}). According to these calculations, a
detection of the gamma ray flux profiles seems to be very hard. We
computed also the prospects for an excess signal detection, i.e. we
are not interested in the shape of the gamma ray flux profile, only in
detectability (Table.~\ref{tab1}). According to the results we reached
there is no chance to detect a $\gamma$-ray signal (flux profiles or
just an excess) coming from Draco with current experiments, at least
with the preferred particle physics and astrophysics models (for a
more detailed study, see the complete work \citep{masc}). It will be
necessary go a step further with IACTs that join a large field of view
with a high sensitivity.

\begin{table}[!h]
\begin{tabular}{ccccc}
\hline
&  & F$_{Draco}$ ($ph~cm^2~s^{-1}$) &  & F$_{min}$ ($ph~cm^2~s^{-1}$) \\
\hline
MAGIC    &  & 1.6~10$^{-19}$ - 4.0~10$^{-16}$ & & 4.4~10$^{-11}$ (250~h, 5$\sigma$)\\
GLAST    &  & 1.6~10$^{-19}$ - 1.6~10$^{-15}$ & & 3.9~10$^{-10}$ (1~yr, 5$\sigma$)\\ 
\hline
\end{tabular}
\caption{Prospects of an excess signal detection for MAGIC and GLAST. For F$_{Draco}$, the most optimistic and pessimistic values are given in the form F$_{Draco,min}$ - F$_{Draco,max}$. F$_{min}$ represents the minimum detectable flux for each instrument. All values refer to the inner 0.5$\grado$ of the dwarf.}
\label{tab1}
\end{table}


\end{document}